\documentclass[11pt]{article}
\usepackage{graphicx}
\usepackage{amsmath}
\usepackage{amssymb}
\usepackage{amsxtra}
\usepackage{mismath}

\usepackage{cite}

\def\institute#1{\gdef\@institute{#1}}
\let\oldmaketitle\maketitle
\renewcommand\maketitle{\oldmaketitle\noindent\@institute}

\begin{document}

\title{Strong primordial inhomogeneities induced by axion-like scalar field}
\author{M.A.~Krasnov$^{1,2}$\footnote{morrowindman1@mail.ru}, M. Yu.~Khlopov$^{3}$, O.~Trivedi$^4$}
\institute{$^1$ National Research Nuclear University “MEPHI”, 115409 Moscow, Russia\\
$^2$ Research Institute of Physics, Southern Federal University, 344090 Rostov-on-Don, Russia\\
$^3$ Virtual Institute of Astroparticle Physics, 75018 Paris, France\\
$^4$ International Centre for Space Sciences and Cosmology, Ahmedabad University, Ahmedabad 380009, India
}

\maketitle

\begin{abstract}
Our goal is to consider Axion-like particle (ALP) model to investigate the behaviour of space-time in the vicinity of the domain wall, induced by axion-like field. Here we present first-step approximation in our analysis and discuss the applicability of thin-shell approximation.
\end{abstract}

\section{Introduction}\label{s:intro}
Axion-like models have become popular candidates for dark matter and present intriguing opportunities for linking particle physics with cosmology. The search for dark matter components has led to investigations into various hypothetical particles beyond the Standard Model (SM). Originally envisioned as an extension of the Peccei-Quinn (PQ) mechanism \cite{peccei1977cp} to address the strong CP problem in Quantum Chromodynamics (QCD), hese
pseudo-Nambu-Goldstone bosons have been proposed in different frame-
works involving quantum gravity \cite{arvanitaki2010string,svrcek2006axions,conlon2006qcd}. ALPs have garnered significant interest as potential dark matter constituents as despite their inherently feeble masses ($m_a \lesssim 1$ keV), non-thermal production mechanisms in the early universe could have yielded a population of ALPs that persists today, potentially constituting the majority of cold dark matter (CDM). The appeal of ALPs as dark matter candidates lies in their ability to circumvent the limitations of the standard freeze-out mechanism, that affect weakly interacting massive particles (WIMPs) in their dark matter candidature. The low mass of ALPs allows them to remain in thermal equilibrium with the bath of particles in the early universe for an extended period and this period evades the issue of WIMPs becoming too sluggish to interact efficiently after freeze-out, leading to an underabundance of relic WIMPs compared to the observed dark matter density \cite{Buschmann:2019icd,Iwazaki:1997bk,gorghetto2023post,Bernal:2021yyb}.
\par Exploring the theoretical background and experimental limitations on ALP properties is crucial for understanding their potential role in cosmology. Recent advancements in directly observing gravitational waves (GWs) are driving significant progress \cite{LIGOScientific:2016aoc,LIGOScientific:2016sjg,Maggiore:1999vm}. Direct GW observations are anticipated to provide important insights into high-energy physics due to their weak interaction with matter, preserving the characteristics of astrophysical or cosmological events \cite{Rubin:2000dq,Khlopov:2004sc,Garriga:2015fdk,Deng:2016vzb,Liu:2022bvr}. Over the past decades, experimental sensitivities for the direct detection of GWs have significantly improved and numerous new GW observatories are planned worldwide and within this framework, it is imperative to explore various potential sources of GWs and determine the extent to which new physics can be inferred from future observations \cite{Grishchuk:1974ny,Starobinsky:1979ty,Witten:1984rs}. Improved experimental sensitivities for GW detection have led to the planning of numerous new GW observatories worldwide. Among the potential cosmological sources of GWs are topological defects like cosmic strings and domain walls that could have formed in the early universe \cite{Kibble:1976sj}. 
Domain walls are sheet-like topological defects that could form in the early universe when a discrete symmetry is spontaneously broken. Given that discrete symmetries are pervasive in high-energy physics beyond the Standard Model (SM), many new physics models predict the formation of domain walls in the early universe. By examining their cosmological evolution, we can derive several constraints on these models, even if their energy scales exceed those probed by laboratory experiments. Many models of ALP fields, for example, have been considered with regards to creation of domain walls in the early universe \cite{Dunsky:2024zdo,Blasi:2023sej}. 
\par Typically the formation of domain walls is considered problematic in cosmology, as their energy density can quickly dominate the total energy density of the universe which one might take as a contradiction to current observational data. While the formation of domain walls is usually considered problematic in cosmology due to their energy density potentially dominating the total energy density of the universe, their instability might prevent this if the discrete symmetry is only approximate and explicitly broken by a small parameter in the theory. In such cases, the collisions and annihilation of domain walls could produce a significant amount of GWs, potentially resulting in a stochastic GW background in the present universe. Observations of relic GWs can offer insights into the early universe and high-energy physics. ALP fields have been shown to contribute to the formation of closed domain walls in scenarios involving an inflationary universe, potentially impacting the nHz stochastic GW background detected by pulsar timing array facilities and early galaxy formation observed by the James Webb Space Telescope \cite{Guo:2023hyp}.

\section{Thin shell approximation}
In this section we will review and discuss the approach of infinitely thin domain wall. 
For extended analysis, please, see \cite{Deng:2016vzb} and \cite{Tanahashi_2015}. 
\par Let us denote wall's surface as $\Sigma$, which is 3-dimensional spacelike or timelike (this would affect $\epsilon$, it would be $+1$ [spacelike] or $-1$ [timelike]) hyper-surface embedded in a 4-dimensional spacetime ($\mathcal{M}, g_{\mu\nu}$). We put subscripts $\pm$ to denote the value on each side of the hypersurface $\Sigma$. Following aforementioned papers:
\begin{align}
    [A]^\pm &\coloneqq A_+ - A_-,\\
    \{A \}^\pm &\coloneqq A_+ + A_-,\\
    \overline{A}&\coloneqq \cfrac{1}{2}\,\{A \}^\pm.
\end{align}
Let $\xi_\mu$ be the normal unit vector to the $\Sigma$, we can define the induced metric $h_{\mu\nu}$ and the extrinsic curvature $K_{\mu\nu}$ as

\begin{align}
    h_{\mu\nu} &\coloneqq g_{\mu\nu} - \epsilon \xi_\mu \xi_\nu,\\
    K_{\mu\nu} &\coloneqq h_\mu^\alpha \nabla_\alpha \xi_\nu = D_\mu \xi_\nu.
\end{align}

\begin{itemize}
    \item First junction condition is:
\begin{equation}\label{1stJunc}
    [h_{\mu\nu}]^\pm=0.
\end{equation}

    \item Second condition is:
    \begin{equation}
        [K_{\mu\nu}]^\pm=8\pi \epsilon \left( -S_{\mu\nu}+\cfrac{1}{2}\,Sh_{\mu\nu} \right).
    \end{equation}

    \item Shell equation of motion:
    \begin{equation}
        S_{\mu\nu} \overline{K}^{\mu\nu} = [T_{\mu\nu}\xi^\mu \xi^\nu]^\pm.
    \end{equation}

    \item  Shell energy conservation:
    \begin{equation}
        D_\mu S^\mu_\nu = -[T_{\mu\alpha}\xi^\mu h^\alpha_\nu]^\pm.
    \end{equation}
\end{itemize}

$S_{\mu\nu}$ is the energy momentum tensor of matter fields on $\Sigma$ and $S=h^{\mu\nu}S_{\mu\nu}$. For pure tension surface $S_{\mu\nu} = -\sigma h_{\mu\nu}$. $T_{\mu\nu}$ is the energy momentum tensor in the region $\mathcal{M}-\{\Sigma\}$.

The paper \cite{Tanahashi_2015} also contains line element in general form:
\begin{equation}
    ds^2=-e^{2\alpha(t,r)}dt^2+e^{2\beta(t,r)}dr^2+R^2(t,r)d\Omega^2.
\end{equation}
They consider the motion of a spherical shell in this spacetime described by
\begin{equation}
    t=t^\xi(\tau),\, \chi=\chi^\xi(\tau),
\end{equation}
where $\tau$ is proper time associated with the shell trajectory in the radial direction. The coordinate components of the radial tangent vector $v^\mu$ is given by
\begin{equation}
    v^\mu = \left( \cfrac{dt^\xi}{d\tau}, \cfrac{d\chi^\xi}{d\tau}, 0, 0\right)=(t_{,\tau},\chi_{,\tau},0,0).
\end{equation}
Further calculations are based on:
\begin{align}
    v_\mu v^\mu&=-1,\\
    \xi_\mu\xi^\mu&=1,\\
    \xi^\mu v_\mu &= 0.
\end{align}

Following \cite{Tanahashi_2015}, second junction yields:
\begin{align}
    \left[\xi^\mu \partial_\mu \ln{R} \right]^\pm = -4\pi \sigma,\\
    \left[ \xi_\mu (v^\mu_{,\tau}+\Gamma^\mu_{\lambda \sigma}v^\lambda v^\sigma)\right]^\pm = -4\pi \sigma 
\end{align}

We set EMT of cosmological fluids in the form:
\begin{equation}
    T^\pm_{\mu\nu}=(\rho_\pm + p_\pm)u_\mu^\pm u_\nu^\pm + p_\pm g_{\mu\nu}^\pm.
\end{equation}
Equation of motion is as follows:
\begin{equation}
    \left\{\xi_\mu (v^\mu_{,\tau}+\Gamma^\mu_{\lambda \sigma}v^\lambda v^\sigma) + 2\xi^\mu \partial_\mu \ln{R} \right\}^\pm = -\cfrac{2}{\sigma} \left[ (p+\rho)(u_\mu \xi^\mu)^2 + p\right]^\pm.
\end{equation}

If we consider the same fluid within and outside the shell, then it is much simpler:
\begin{equation}
    \left\{\xi_\mu (v^\mu_{,\tau}+\Gamma^\mu_{\lambda \sigma}v^\lambda v^\sigma) + 2\xi^\mu \partial_\mu \ln{R} \right\}^\pm = 0,
\end{equation}
if one assume the smooth crossing of the wall by fluid ($[u_\mu \xi^\mu]^\pm=0$).

Equation of motion of the thin shell could be written as \cite{Deng:2016vzb}:
\begin{equation}
    \cfrac{ar_{,\tau\tau}}{\sqrt{1+a^2r^2_{\tau}}}+\cfrac{4a_{,\tau}r_{,\tau}}{\sqrt{1+a^2r^2_{\tau}}}+\cfrac{2\sqrt{1+a^2r^2_{\tau}}}{ar}=6\pi\sigma,
\end{equation}
which can also be rewritten in terms of cosmic time as follows:
\begin{equation}\label{ShellEqofmotion}
    \Ddot{r}+(4-3a^2\Dot{r}^2)H\Dot{r}+\cfrac{2}{a^2r}\,(1-a^2\Dot{r}^2)=6\pi\sigma\,\cfrac{(1-a^2\Dot{r}^2)^{3/2}}{a}.
\end{equation}
One may notice that equation of motion \eqref{ShellEqofmotion} could be utilized in two cases with regards to wall's energy density. The first case is wall's energy density is negligible, i.e. one can put $\sigma = 0$. And the second case is wall's density is considerable, but not arbitrarily big, since there is no corresponding transformation for equation \eqref{ShellEqofmotion}, i.e. there is no limit of big $\sigma$, which makes it inapplicable for consideration of wall-dominated Universe in this framework.

\section{Axion-like field in Kantowski-Sachs metric}

 We start by considering a space-time with line element given by 
\begin{equation}\label{InitialLineEl}
    ds^2=A(t,r)^2dt^2 - X(t,r)^2dr^2-Y(t,r)^2d\Omega^2,
\end{equation}
where $d\Omega^2 = d\theta^2+\sin^2{\theta}d\phi^2$. One may notice abuse of symbols in our paper, we denote axion-like scalar field with both $\theta$ and $\phi$ symbols as well as angular coordinates. But there is no explicit utilization of angular coordinates in our analysis, thus, there would be no misunderstanding.  
Following \cite{Deng:2016vzb}, we set $A(t,r) = 1$, corresponding to gauge symmetry.
ALP Lagrangian is set to be
\begin{equation}\label{lagr}
    \mathcal{L} = \cfrac{1}{2}\,g^{\mu\nu}\partial_\mu\phi \partial_\nu\phi - \Lambda^4 \left[1-\cos{\left(\frac{\phi}{f} \right)}\right]. 
\end{equation}
We also take into account anisotropy of the space time via energy-momentum tensor of cosmological fluid:
 {
\begin{equation}\label{EMT}
     T^\mu_\nu=\text{diag}[1, \omega, -(\omega+\delta), -(\omega+\gamma)]\rho,
\end{equation}
where $\omega=p/\rho$.
}
 
 {As a first step in our analysis we consider limit $r\rightarrow \infty$, i.e. what a distant observer could detect. In that case line element \eqref{InitialLineEl} could be rewritten as
\begin{equation}\label{KSmetric}
    ds^2=dt^2 - X(t)^2dr^2-Y(t)^2d\Omega^2,
\end{equation}
which is Kantowski-Sachs space time \cite{kantowski1966some}. Here we assume that $Y$ is finite as $r\rightarrow \infty$. We also set common border condition for the scalar field $\phi_r(t, \infty) = 0.$ 
\\
\\
Now let us write down system of Einstein's field equations using line element \eqref{KSmetric}, Lagrangian \eqref{lagr} and cosmological fluid' energy-momentum tensor \eqref{EMT}:}

\begin{align}
    \frac{2 \Dot{X} \Dot{Y}}{X Y}+\frac{\Dot{Y}^2}{Y^2}+\frac{1}{Y^2} &= \cfrac{1}{2}\,\Dot{\phi}^2+\Lambda^4\left[ 1-\cos{\left(\cfrac{\phi}{f}\right)}\right] + \rho, \label{eq1}\\
    \frac{2 \Ddot{Y}}{Y}+\frac{\Dot{Y}^2}{Y^2}+\frac{1}{Y^2} &= -\cfrac{1}{2}\,\Dot{\phi}^2+\Lambda^4\left[ 1-\cos{\left(\cfrac{\phi}{f}\right)}\right] -\omega\rho, \label{eq2}\\
    \frac{\Ddot{X}}{X}+\frac{\Dot{X} \Dot{Y}}{X Y}+\frac{\Ddot{Y}}{Y} &= -\cfrac{1}{2}\,\Dot{\phi}^2+\Lambda^4\left[ 1-\cos{\left(\cfrac{\phi}{f}\right)}\right] -(\omega+\delta)\rho, \label{eq3}\\
    \frac{\Ddot{X}}{X}+\frac{\Dot{X} \Dot{Y}}{X Y}+\frac{\Ddot{Y}}{Y} &= -\cfrac{1}{2}\,\Dot{\phi}^2+\Lambda^4\left[ 1-\cos{\left(\cfrac{\phi}{f}\right)}\right] -(\omega+\gamma)\rho. \label{eq4}
\end{align}
Equation of motion of the scalar field is as follows:
\begin{equation}\label{kgeq}
    \Ddot{\phi}+\left( \cfrac{\Dot{X}}{X}+2\,\cfrac{\Dot{Y}}{Y}\right)\Dot{\phi}+\cfrac{\Lambda^4}{f}\,\sin{\left( \cfrac{\phi}{f} \right)}=0.
\end{equation}
 {Equations \eqref{eq3} and \eqref{eq4} immediately yield $\gamma = \delta$ and now we are given with four independent equations and six variables. We need to make two assumptions to make system solvable.}
\\
\\
To make progress now, we assume the following relation between $X$ and $Y$: 
\begin{equation}\label{AnisotSol}
    Y=X^n,
\end{equation}
 {which has been commonly utilized in some previous works by other researchers \cite{Collins, LRS}.} Furthermore, it is usually assumed that scalar field is proportional to average scale factor taken in some power \cite{johri1994cosmological, johri1989bd} 
\begin{equation}\label{fieldandscale}
    \phi \propto a(t)^l = (XY^2)^{l/3},
\end{equation}

\begin{figure}[ht]\label{imp_of_l}
	\begin{center}
\includegraphics[width=1\textwidth]{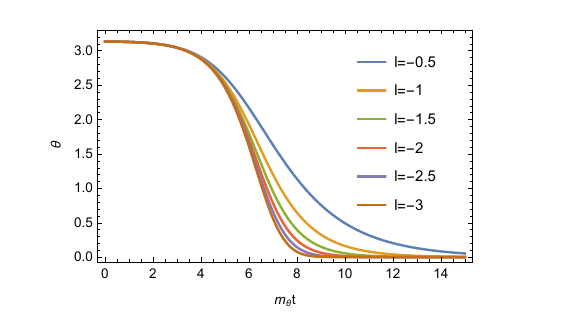}
	\end{center}
\caption{The plot of the solution of \eqref{thetaEq} for different values of of $l$, where the interesting observation is the impact of $l$ changing slowly for $l<-2$.}	
\end{figure}

 {One can then check that} relations \eqref{fieldandscale} and  \eqref{AnisotSol} leads to the following relation between the scalar field and metric's potential
\begin{equation}\label{assumption}
    \cfrac{\Dot{\phi}}{\phi}=\cfrac{l}{3}\,(2n+1)\,\cfrac{\Dot{X}}{X}.
\end{equation}
We now utilize \eqref{assumption} in \eqref{kgeq}, which leads us to arrive at
\begin{equation}\label{maineq}
    \Ddot{\phi}+\cfrac{3\Dot{\phi}^2}{l\phi}+\cfrac{\Lambda^4}{f}\,\sin{\left( \cfrac{\phi}{f} \right)}=0.
\end{equation}
 {Let us now perform the following variable substitution $\phi = f\theta$ and switch from time derivative to the derivative with respect to $m_\theta t=t\Lambda^2/f$ (represented by prime). We obtain} 

\begin{equation}\label{thetaEq}
    \theta'' + \cfrac{3}{l}\,\cfrac{\theta'^2}{\theta}+\sin{\theta}=0.
\end{equation}
At this point, we could refer to $\theta$ as a phase and we could then plot the numerical solution of \eqref{thetaEq} for different values of $l$. We set initial conditions for $\theta$ as follows
\begin{equation}
    \theta_{in}=\pi-0.01,\,\,\theta'_{in}=0.
\end{equation}
 {This allows us to see the evolution of the scalar field, which we have shown in Figure \ref{imp_of_l}.}

\begin{figure}[ht]
	\begin{center}
\includegraphics[width=1\textwidth]{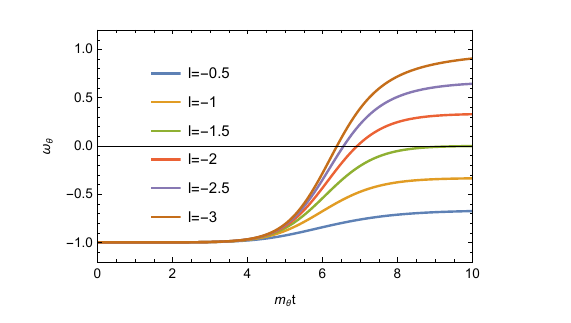}
	\end{center}
\caption{ {The plot of the equation of state parameter for the scalar field. We see that scalar field could behave differently depending on the value of $l$ and in particular, if $l=-1.5$ then scalar field behaves like non-relativistic matter.}}	\label{imp_of_l_omega_theta}
\end{figure}

 {The equation of state parameter, given by the usual definition $$\omega_\theta = \frac{p_\theta}{\rho_\theta}$$ is plotted in Fig.\eqref{imp_of_l_omega_theta}} \footnote{ {Note that the parameter has also been scaled appropriately with regards to the differential equation with $\theta$ variable.}}

 {From the Einstein equations \eqref{eq1}-\eqref{eq4}, we find that equation of state parameter for cosmological fluid $\omega \approx \text{const} \sim-1$ and skewness parameter $\delta=\gamma \approx 1$ for any value of $l$.}

\section{Conclusion}
In this study, we explored the cosmological behavior of a scalar field with an axion-like potential in a non-standard spacetime background. The metric we examined was generally that of an inhomogeneous and anisotropic spacetime, which, in the limit of a distant observer, resembled a standard Kantowski-Sachs metric. Specifically focusing on this scenario, we analyzed the dynamics of the scalar field in such a setting.

In the model for a distant observer, the parameters of skewness, represented by $ \delta $ and $ \gamma $, tended towards minus one, resulting in negligible angular pressures and causing the energy density of the cosmological fluid to align radially. This anisotropic nature suggests that the fluid could support cosmic expansion in specific directions while exhibiting dust-like behavior in angular dimensions.

\section*{Acknowledgements}
We are grateful to Yu.N. Eroshenko for fruitful discussions. 


\end{document}